# Mass Splitting and Production of $\Sigma_c^0$ and $\Sigma_c^{++}$ Measured in 500 GeV $\pi^-$–N Interactions

E. M. Aitala,[8] S. Amato,[1] J. C. Anjos,[1] J. A. Appel,[5] D. Ashery,[14] S. Banerjee,[5]
I. Bediaga,[1] G. Blaylock,[2] S. B. Bracker,[15] P. R. Burchat,[13] R. A. Burnstein,[6]
T. Carter,[5] H. S. Carvalho,[1] N. K. Copty,[12] I. Costa,[1] L. M. Cremaldi,[8]
C. Darling,[18] K. Denisenko,[5] J. Engelfried,[5] A. Fernandez,[11] P. Gagnon,[2]
S. Gerzon,[14] K. Gounder,[8] A. M. Halling,[5] G. Herrera,[4] G. Hurvits,[14] C. James,[5]
P. A. Kasper,[6] N. Kondakis,[10] S. Kwan,[5] D. C. Langs,[10] J. Leslie,[2] J. Lichtenstadt,[14]
B. Lundberg,[5] A. Manacero,[5] S. MayTal-Beck,[14] B. Meadows,[3]
J. R. T. de Mello Neto,[1] R. H. Milburn,[16] J. M. de Miranda,[1] A. Napier,[16]
A. Nguyen,[7] A. B. d'Oliveira,[3,11] K. O'Shaughnessy,[2] K. C. Peng,[6] L. P. Perera,[3]
M. V. Purohit,[12] B. Quinn,[8] S. Radeztsky,[17] A. Rafatian,[8] N. W. Reay,[7]
J. J. Reidy,[8] A. C. dos Reis,[1] H. A. Rubin,[6] A. K. S. Santha,[3] A. F. S. Santoro,[1]
A. J. Schwartz,[10] M. Sheaff,[17] R. A. Sidwell,[7] A. J. Slaughter,[18] J. G. Smith,[7]
M. D. Sokoloff,[3] N. R. Stanton,[7] K. Sugano,[2] D. J. Summers,[8] S. Takach,[18]
K. Thorne,[5] A. K. Tripathi,[9] S. Watanabe,[17] R. Weiss-Babai,[14] J. Wiener,[10]
N. Witchey,[7] E. Wolin,[18] D. Yi,[8] R. Zaliznyak,[13] C. Zhang[7]
(Fermilab E791 Collaboration)

[1] *Centro Brasileiro de Pesquisas Físicas, Rio de Janeiro, Brazil*
[2] *University of California, Santa Cruz, California 95064*
[3] *University of Cincinnati, Cincinnati, Ohio 45221*
[4] *CINVESTAV, Mexico City, Mexico*
[5] *Fermilab, Batavia, Illinois 60510*
[6] *Illinois Institute of Technology, Chicago, Illinois 60616*
[7] *Kansas State University, Manhattan, Kansas 66506*
[8] *University of Mississippi, University, Mississippi 38677*
[9] *The Ohio State University, Columbus, Ohio 43210*
[10] *Princeton University, Princeton, New Jersey 08544*
[11] *Universidad Autonoma de Puebla, Puebla, Mexico*

[12] *University of South Carolina, Columbia, South Carolina 29208*

[13] *Stanford University, Stanford, California 94305*

[14] *Tel Aviv University, Tel Aviv, Israel*

[15] *317 Belsize Drive, Toronto, Canada*

[16] *Tufts University, Medford, Massachusetts 02155*

[17] *University of Wisconsin, Madison, Wisconsin 53706*

[18] *Yale University, New Haven, Connecticut 06511*

## Abstract

From a sample of $2722 \pm 78$ $\Lambda_c^+$ decaying to the $pK^-\pi^+$ final state, we have observed, in the hadroproduction experiment E791 at Fermilab, $143 \pm 20$ $\Sigma_c^0$ and $122 \pm 18$ $\Sigma_c^{++}$ through their decays to $\Lambda_c^+\pi^\pm$. The mass difference $M(\Sigma_c^0) - M(\Lambda_c^+)$ is measured to be $(167.38 \pm 0.29 \pm 0.15)$ MeV; for $M(\Sigma_c^{++}) - M(\Lambda_c^+)$, we find $(167.76 \pm 0.29 \pm 0.15)$ MeV. The rate of $\Lambda_c^+$ production from decays of the $\Sigma_c$ triplet is $(22 \pm 2 \pm 3)$ % of the total $\Lambda_c^+$ production assuming equal rate of production from all three, as measured for $\Sigma_c^0$ and $\Sigma_c^{++}$. We do not observe a statistically significant $\Sigma_c$ baryon-antibaryon production asymmetry. The $x_F$ and $p_t^2$ spectra of $\Lambda_c^+$ from $\Sigma_c$ decays are observed to be similar to those for all $\Lambda_c^+$'s produced.

Mass splittings among baryons containing different valence quarks provide a good probe of hadronic structure and of the forces that determine this structure. Isospin mass splittings have been attributed to the intrinsic difference between the light quark masses, and to the electromagnetic interactions among the quarks [1]. These splittings have been calculated with quark models and sum rules based on QCD [2–11]; in some of these calculations, it has been suggested that the color hyperfine mass splittings from the spin-spin interactions may also be important [12]. Experimentally, these mass splittings can be measured and different dynamical models can be tested. Especially interesting are the mass splittings in the charm baryon system, because the simplifications expected due to the large mass of the $c$ quark allow more meaningful tests of models to be made. A disagreement of the $\Sigma_c$ mass splitting measurements with sum rule predictions has recently been cited as posing a significant problem for the charm baryon sector of the quark model [13].

Measurement of charm baryon mass differences is most accessible through the hadronic decay of the $I = 1$ partner of the $\Lambda_c^+$, $\Sigma_c \to \Lambda_c^+\pi$. Throughout this paper, the charge conjugate state is implicitly included. The $\Sigma_c$'s are expected to be copiously produced in the hadroproduction of charm particles, and studies of the production characteristics of these charm baryon states provide information about the process of heavy quark hadronization. But to date, the only fixed-target hadroproduction experiment that has published results on $\Sigma_c$ production is NA32 (ACCMOR) [14], based on a sample of 150 $\Lambda_c^+ \to pK^-\pi^+$ candidates. They reported that from 230 GeV $\pi^- - N$ interactions, less than 5.3 % (90 % C.L.) of $\Lambda_c^+$ decays come from $\Sigma_c^0$ and less than 5.2 % (90 % C.L.) from $\Sigma_c^{++}$ decays. This can be compared to $e^+e^-$ [15–17] and photoproduction experiments [18,19] which report that the $\Sigma_c$ triplet accounts for about 25 % of the total $\Lambda_c^+$ production.

In this paper, we report the observation of two of the three isospin states of the $\Sigma_c$, the $\Sigma_c^0$ and the $\Sigma_c^{++}$, and a measurement of the mass difference between each of these two states and the $\Lambda_c^+$. The results are based on $2722 \pm 78$ $\Lambda_c^+ \to pK^-\pi^+$ candidates produced in 500 GeV $\pi^- - N$ interactions in experiment E791 at Fermilab. Previous mass difference measurements have either been of low statistics or used several decay modes of the $\Lambda_c^+$ added together [15–17,19], with different systematic uncertainties for each mode. We report measurements of the rate of $\Lambda_c^+$ production from $\Sigma_c$ decays relative to the inclusive rate for $\Lambda_c^+$. We also determine this relative rate as a function of Feynman-$x$ ($x_F \equiv p_z/p_z^{max}$) and transverse momentum squared ($p_t^2$).

The results given in this paper are based on the full dataset accumulated by experiment E791 in the 1991/92 Fermilab fixed-target run. The segmented target used in the experiment consisted of a 0.5 mm thick platinum foil followed by four diamond foils, each 1.6 mm thick. Each target center was separated from the next one by about 1.5 cm, thus allowing the observation of the decay of the $\Lambda_c^+$ in air, minimizing the background from secondary interactions. The spectrometer [20] used in E791 was essentially the same as that used in experiments E691 and E769. It is a large-acceptance two-magnet spectrometer with eight planes of multiwire proportional chambers (MWPC) and six planes of silicon microstrip detectors (SMD) before the target for beam tracking, a 17-plane SMD system downstream of the target for track and vertex reconstruction, 35 drift chamber planes, two MWPC's, two multi-cell threshold Cherenkov counters, electromagnetic and hadronic calorimeters, and a muon detector. An important element of the experiment was its extremely fast data acquisition system [21] which was combined with a very open transverse-energy trigger to record a data sample of 20 billion interactions.

The analysis begins with the selection of $\Lambda_c^+ \to pK^-\pi^+$ candidate decays. For this, all events were first reconstructed and a candidate vertex position for the primary interaction was determined. Events were selected for further analysis if at least one secondary vertex was also reconstructed. Tracks positively identified by the Cherenkov counters as kaons and protons were combined with the remaining tracks (assumed to be pions) in the event to search for $\Lambda_c^+ \to pK^-\pi^+$ candidate decays.

The final $\Lambda_c^+$ sample was selected by the following requirements. The significance of separation of the candidate decay vertex from the primary vertex in the longitudinal direction was required to be greater than $7\,\sigma_\Delta$ where $\sigma_\Delta$ is the measurement error on the longitudinal separation of the two vertices. The decay track candidates were each required to be inconsistent with having originated in the primary vertex by requiring the transverse separation from the primary vertex to be greater than $3\,\sigma_t$. The distance of closest approach between the line of flight of the reconstructed decaying particle and the primary vertex was required to be less than $40\,\mu$m. The scalar sum of $p_t^2$ of the decay tracks, with $p_t$ measured relative to the direction of the parent, was required to be greater than $0.25\,\mathrm{GeV}^2$. We calculated, for each decay track, the ratio of its impact parameters to the secondary and primary vertices. The product of these ratios from the three decay tracks was required to be less than 0.01. Finally, the decay vertex had to be located at least $4\,\sigma_z$ (where $\sigma_z$ is the error on the position of the decay vertex along the beam direction) outside the target foils to remove secondary interactions. The final $pK^-\pi^+$ invariant mass distribution is shown in Fig. 1. From a fit to a Gaussian shape for the signal and a linear term for the background, a $\Lambda_c^+$ signal of $2722 \pm 78$ events was obtained. Due to the low Cherenkov misidentification rate and the good mass resolution of the spectrometer, the contamination from misidentified $D^+$ and $D_s^+$ decays in the $\Lambda_c^+$ signal is less than 5 %.

To search for the decay $\Sigma_c \to \Lambda_c^+\pi$, we add a pion of either charge to the $pK^-\pi^+$ state. This "bachelor" pion is required to be inconsistent with a kaon or proton according to information from the Cherenkov counters and to be not associated with another decay vertex reconstructed in the event. The mass differences between the $\Sigma_c$ and $\Lambda_c^+$ candidates, $\Delta M = M(pK^-\pi^+\pi^\pm) - M(pK^-\pi^+)$, are shown in Figs. 2 and 3 for $\Sigma_c^0$ and $\Sigma_c^{++}$, respectively, for $pK^-\pi^+$ combinations within $2.5\,\sigma_M$ of the fitted $\Lambda_c^+$ mass, where $\sigma_M$ is the measured width of the $\Lambda_c^+$ signal (10 MeV).

The mass difference spectra are fit to Gaussian distributions of fixed width and background shapes. The latter are relativistic phase space approximations, proportional to $[1 + \alpha(\Delta M - m_\pi)\Delta M^\beta]$ where $\alpha$ and $\beta$ are free parameters in the fits. The width of the Gaussian signal term is determined by Monte Carlo studies [22] of the experimental resolution and fixed at 2.0 MeV for both the $\Sigma_c^0$ and $\Sigma_c^{++}$. The result of an unbinned maximum likelihood fit to the mass difference plot for candidate $\Sigma_c^0$ decays (Fig. 2) yields $143 \pm 20$ $\Sigma_c^0$ events with a mean $\Delta M$ of $(167.44 \pm 0.29)$ MeV. For Fig. 3, the fit gives $122 \pm 18$ $\Sigma_c^{++}$ events with a mass difference of $(167.82 \pm 0.29)$ MeV.

The natural widths of the $\Sigma_c^0$ and $\Sigma_c^{++}$ resonances [12] are predicted to be small compared to the experimental resolution and are not included in the fit. Increasing or reducing the width of the Gaussian by 0.5 MeV changes the central value of $\Delta M$ by less than 0.1 MeV.

We have studied the systematic uncertainty due to the background determination by three methods: 1) using different background parametrizations, 2) fixing the constants $\alpha$ and $\beta$ to values determined by fitting the background events in the wings of the $\Lambda_c^+$ signal, and 3) using random combinatoric background given by a $\Lambda_c^+$ candidate from one event and a "bachelor" pion candidate from another event to determine $\alpha$ and $\beta$. We found that the central value of the peak varies by less than 0.03 MeV. Since we measure the mass differences between the $\Sigma_c$ and the $\Lambda_c^+$ baryons instead of measuring the $\Sigma_c$ mass directly, most of the experimental uncertainties from the reconstruction of the $\Lambda_c^+$ cancel. The remaining systematic uncertainty comes from the measurement of the momentum and decay angle of the "bachelor" pion. We use our measurement of the mass difference $M(D^{*+}) - M(D^0)$ from $D^{*+} \to D^0\pi^+$ decays to estimate this systematic error, since the momentum spectrum of the bachelor pions is similar and the mass difference technique is identical. Our measured mass difference $M(D^{*+}) - M(D^0)$ is $(145.47 \pm 0.01)$ MeV to be compared to the PDG fitted value [23] of $(145.42 \pm 0.05)$ MeV. Using our Monte Carlo, we find that the reconstructed mass difference for the $D^{*+} - D^0$ is shifted by $(+0.05 \pm 0.01)$ MeV relative to the generated mass difference. Similarly, comparing the reconstructed and generated values, the average shift in the mass difference between the $\Sigma_c$'s and $\Lambda_c^+$ is $(+0.06 \pm 0.04)$ MeV. We use $\phi \to K^+K^-$ decays (which have a similar $Q$ to $\Sigma_c$ decays) to estimate the overall uncertainty in our mass difference scale to be $\pm 0.11$ MeV. Finally, we estimate a systematic uncertainty of 0.07 MeV from the observed spread of the measured mass differences when the analysis cuts are varied.

After correcting for the shift in the mass according to the Monte Carlo, our final result for the mass difference between $\Sigma_c^0$ and $\Lambda_c^+$ is $(167.38 \pm 0.29 \pm 0.15)$ MeV; for $M(\Sigma_c^{++}) - M(\Lambda_c^+)$, the final result is $(167.76 \pm 0.29 \pm 0.15)$ MeV. The corresponding values from the PDG fit are $(167.3 \pm 0.4)$ MeV and $(168.04 \pm 0.27)$ MeV. The only recent results on these mass differences come from a photoproduction experiment which measured $M(\Sigma_c^0)-M(\Lambda_c^+)$ to be $(166.6\pm0.5\pm0.6)$ MeV and $M(\Sigma_c^{++})-M(\Lambda_c^+)$ to be $(167.6\pm0.6\pm0.6)$ MeV [19]. Our results provide the best single measurements of these mass differences to date. We find $\Delta M$ $(\Sigma_c^{++}-\Sigma_c^0)$ to be $(0.38\pm0.40\pm0.15)$ MeV which is in good agreement with the latest PDG value of $(0.7\pm0.4)$ MeV. Theoretical predictions for this mass difference range from $-18.0$ to $+6.5$ MeV and are listed in Table 1. The best agreement with data is obtained by Chan [4] who assumed in the framework of an SU(8) model, that hadron mass differences result from intrinsic quark mass differences and two-body spin-spin interactions together with the Coulomb and magnetic interactions.

To calculate the rate of $\Lambda_c^+$ production from $\Sigma_c$ decays relative to that for total inclusive $\Lambda_c^+$ production, we use the Monte Carlo simulation to determine the reconstruction efficiencies of $\Lambda_c^+$, $\Sigma_c^0$ and $\Sigma_c^{++}$. Our acceptance covers the kinematic region in the $x_F$ range above $-0.1$. Within this $x_F$ range, the ratio of the efficiency for reconstructing $\Sigma_c^{0,++}\to\Lambda_c^+\pi^{-,+}$, $\Lambda_c^+\to pK^-\pi^+$ to the reconstruction efficiency for inclusive $\Lambda_c^+\to pK^-\pi^+$ decays is $(66.5\pm1.4)$ % and is the same for particle and antiparticle, and for both $\Sigma_c$ charge states. Furthermore, this ratio of acceptances is found to be independent of $x_F$ within statistical errors so that our result is independent of the shape of the generated $x_F$ distribution of the $\Lambda_c^+$. Assuming the branching ratio $Br(\Sigma_c\to\Lambda_c^+\pi)=1$, we find that $(7.9\pm1.1\pm1.0)$ % of $\Lambda_c^+$ comes from $\Sigma_c^0$ and $(6.7\pm1.0\pm1.0)$ % from $\Sigma_c^{++}$ decays. The first error is statistical and the second one is systematic. The estimate of the systematic error comes from varying the width used in the fit for the mass difference between the $\Sigma_c$ and the $\Lambda_c^+$, using different background parametrizations, varying the analysis cuts and from the uncertainty in the tracking efficiency.

If we assume equal production of states within the $\Sigma_c$ isotriplet and average over the observed rate of production of $\Sigma_c^0$ and $\Sigma_c^{++}$, then $\sigma_{\Sigma_c}/\sigma_{\Lambda_c^+}=(7.3\pm0.8\pm1.0)$ % for each mode, or $(22\pm2\pm3)$ % for all $\Sigma_c$ production relative to $\Lambda_c^+$ production. This is significantly higher than the previous upper limits reported for a 230 GeV $\pi^-$ beam [14]. Our 500 GeV result is consistent with the 25 % seen in photoproduc-tion [18,19] and $e^+e^-$ experiments [15,16] (Table 2). We have also checked for $\Sigma_c$ baryon-antibaryon production asymmetry. In $\pi^- - N$ interactions, an enhanced production of leading $D^-$ mesons has been observed compared to that for non-leading $D^+$ mesons in the forward region [24–26]. We measure a production ratio for $\Sigma_c^0/\overline{\Sigma}_c^0$ of $1.34 \pm 0.39$ and a production ratio for $\Sigma_c^{++}/\overline{\Sigma}_c^{--}$ of $0.98 \pm 0.30$. We do not observe a statistically significant leading particle effect.

To measure the relative production rate as a function of $x_F$, we extract the number of $\Lambda_c^+$ decays in six bins of $x_F$ in the range $-0.1$ to 0.5. We then impose a $\pm 2.5\,\sigma_M$ cut on the $pK^-\pi^+$ mass and plot the difference between $M(pK^-\pi^+\pi^\pm)$ and $M(pK^-\pi^+)$ for the same range of $x_F$. Each distribution is then fit to give the number of $\Sigma_c$ decays. Similarly, we divide the signal into different $p_t^2$ bins between 0 and $10\,\mathrm{GeV}^2$. We have used the Monte Carlo to check the ratios of acceptances between all $\Lambda_c^+$ and those which come from $\Sigma_c$ decays as functions of $x_F$ and $p_t^2$. Within statistics, both ratios of the two acceptances are flat. After correcting for acceptances, we check the rate of production of $\Lambda_c^+$ from $\Sigma_c$ decays relative to the production for all $\Lambda_c^+$ as a function of $x_F$ and $p_t^2$. The relative rate is found to be constant, showing no dependence on $x_F$ or $p_t^2$. This implies that the $x_F$ and $p_t^2$ distributions for $\Lambda_c^+$ from $\Sigma_c$ decays are similar to those from all $\Lambda_c^+$, and subsequently that the $\Sigma_c$ production spectrum is similar to the $\Lambda_c$ spectrum (since the decay $\pi$ is from a low $Q$ decay).

In summary, we report the first observation of $\Sigma_c^0$ and $\Sigma_c^{++}$ in $\pi^- - N$ interactions. We measure the difference between the $\Sigma_c^0$ and $\Lambda_c^+$ masses to be $(167.38 \pm 0.29 \pm 0.15)\,\mathrm{MeV}$ and the difference between $\Sigma_c^{++}$ and $\Lambda_c^+$ massses to be $(167.76 \pm 0.29 \pm 0.15)\,\mathrm{MeV}$. We observe an isospin mass splitting between the $\Sigma_c^{++}$ and $\Sigma_c^0$ states of $(0.38 \pm 0.40 \pm 0.15)\,\mathrm{MeV}$. We determine the rate of $\Lambda_c^+$ production from the decay of the $\Sigma_c$ triplet assuming isospin symmetry to be $(22 \pm 2 \pm 3)\,\%$ of the total $\Lambda_c^+$ production in 500 GeV $\pi^-$–N interactions.

We gratefully acknowledge the assistance of the staffs of Fermilab and of all the participating institutions. This research was supported by the Brazilian Conselho Nacional de Desenvolvimento Científico e Tecnológico, CONACyT (Mexico), the Israeli Academy of Sciences and Humanities, the U.S. Department of Energy, the U.S.-Israel

Binational Science Foundation and the U.S. National Science Foundation. Fermilab is operated by the Universities Research Association, Inc., under contract with the United States Department of Energy.

## TABLES

| Reference | $\Delta M(\Sigma_c^{++} - \Sigma_c^0)$ in MeV |
|---|---|
| Capstick [1] | 1.4 |
| Lane [3] | −6.0 |
| Chan [4] | 0.4 |
| Hwang [5] | 3.0 |
| Sinha [6] | $1.5 \pm 0.2$ |
| Itoh [7] | 6.5 |
| Kalman [8] | −2.7 |
| Lichtenberg [9] | 3.4 |
| Deshpande [10] | −3.0 to −18.0 |
| Richard [11] | −3.0 or 2.0 |
| Kwong [12] | 2.6 |

TABLE I. Theoretical calculations for $\Delta M(\Sigma_c^{++} - \Sigma_c^0)$. These predictions can be compared to our result of $(0.38 \pm 0.40 \pm 0.15)$ MeV.

| Baryon | $e^+e^-$ [16] | 145 GeV $\gamma$ [18] | 220 GeV $\gamma$ [19] | 230 GeV $\pi^-$ [14] | 500 GeV $\pi^-$ (our result) |
|---|---|---|---|---|---|
| $\Sigma_c^0$ | $(9 \pm 3)\,\%$ | $(13 \pm 4 \pm 2)\,\%$ | $(7.8 \pm 2.1)\,\%$ | $< 5.3\,\%$(90 %C.L.) | $(7.9 \pm 1.5)\,\%$ |
| $\Sigma_c^{++}$ | $(9 \pm 3)\,\%$ | $(5 \pm 3 \pm 2)\,\%$ | $(6.7 \pm 1.9)\,\%$ | $< 5.2\,\%$(90 %C.L.) | $(6.7 \pm 1.4)\,\%$ |

TABLE II. Rate of $\Sigma_c$ production relative to total inclusive $\Lambda_c$ production for different beams.

## FIGURES

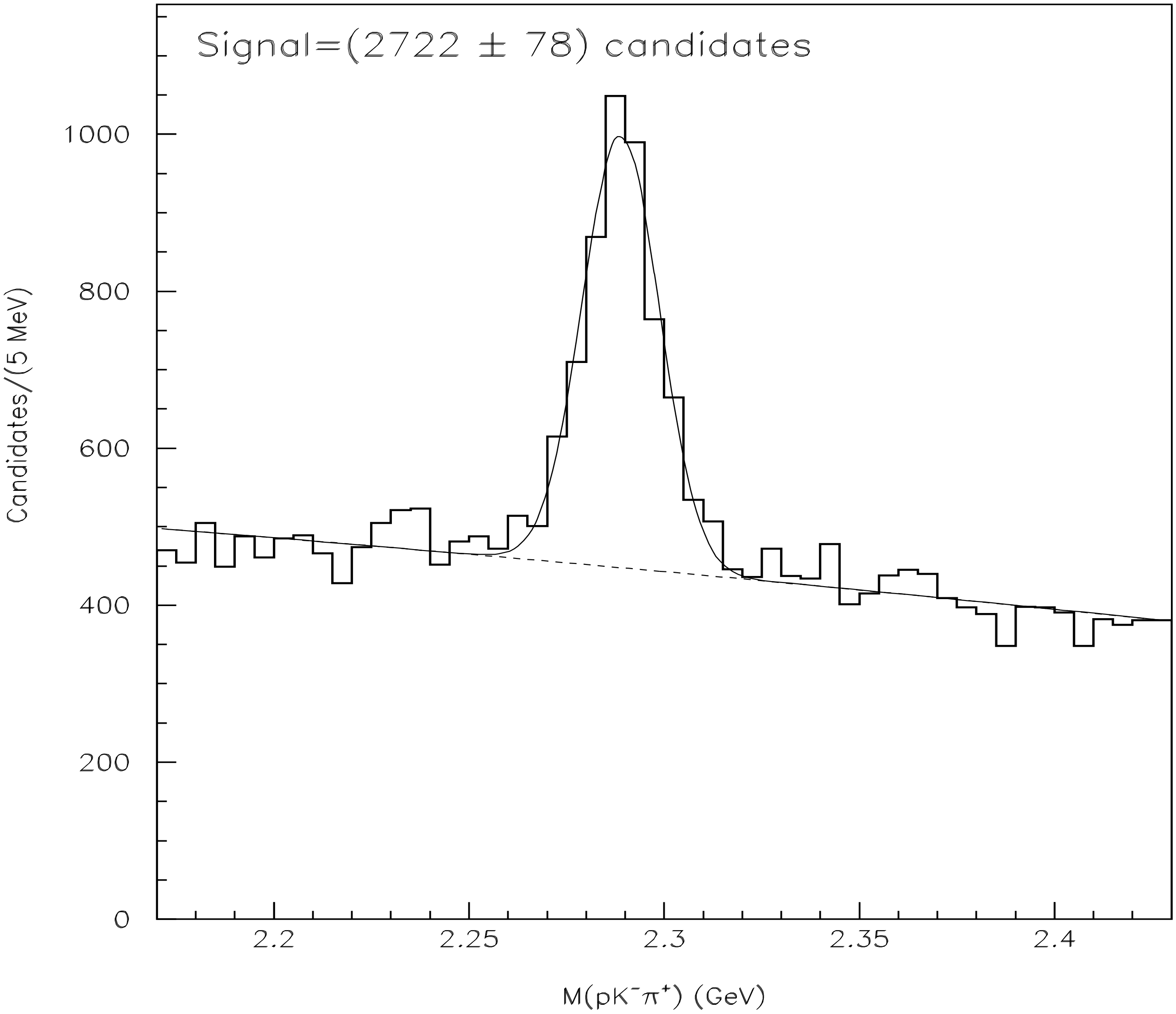

FIG. 1. Invariant mass distribution for candidate $\Lambda_c^+ \to pK^-\pi^+$ decays. The solid line corresponds to a fit to a Gaussian distribution plus a linear background. The dashed line shows the background alone.

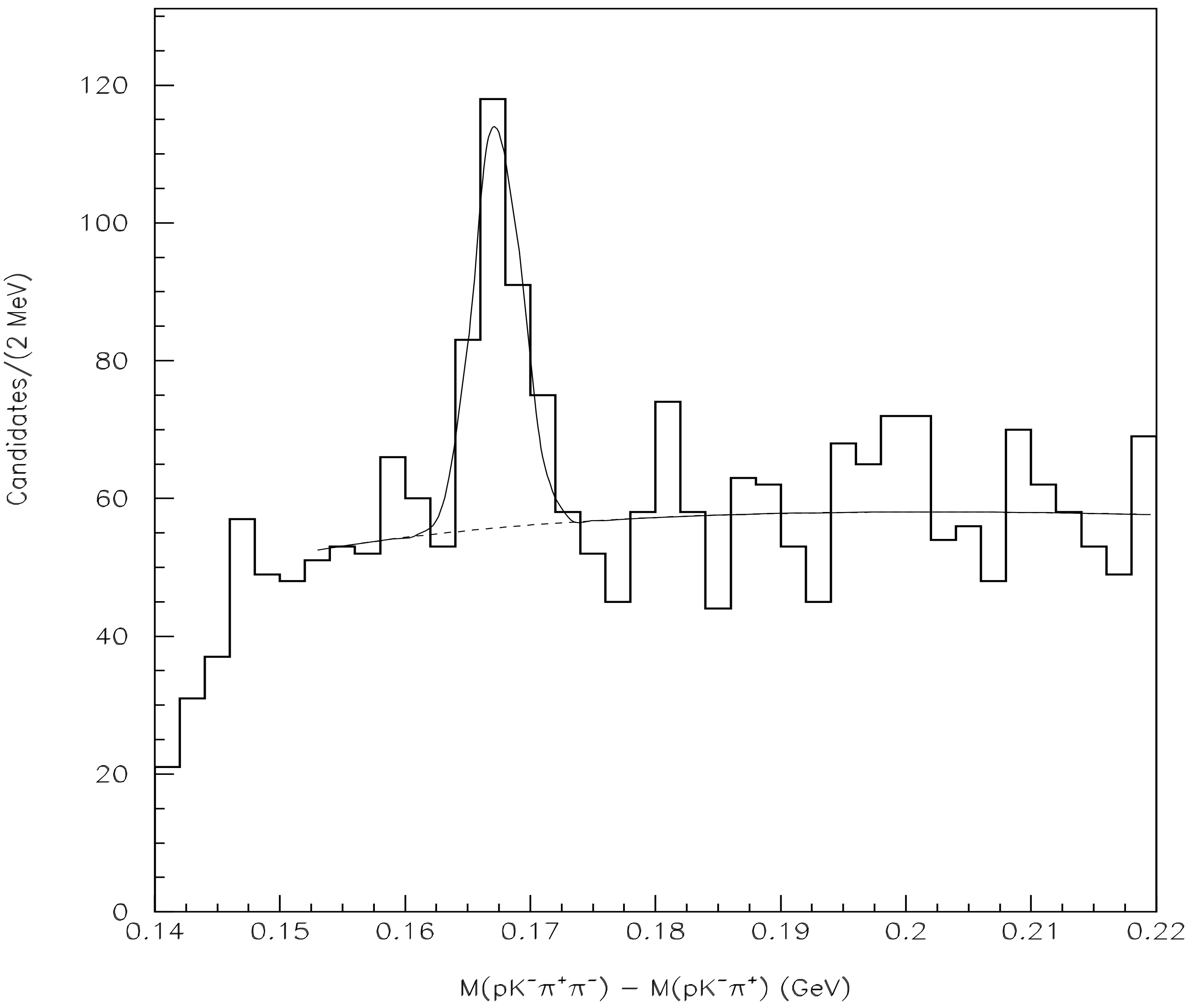

FIG. 2. Mass difference spectrum, M($pK^-\pi^+\pi^-$) − M($pK^-\pi^+$), for $\Sigma_c^0 \to pK^-\pi^+\pi^-$ candidates with $pK^-\pi^+$ combinations within 2.5 $\sigma$ of the fitted $\Lambda_c$ mass. The solid line is a fit to a Gaussian distribution plus a background shape described in the text.

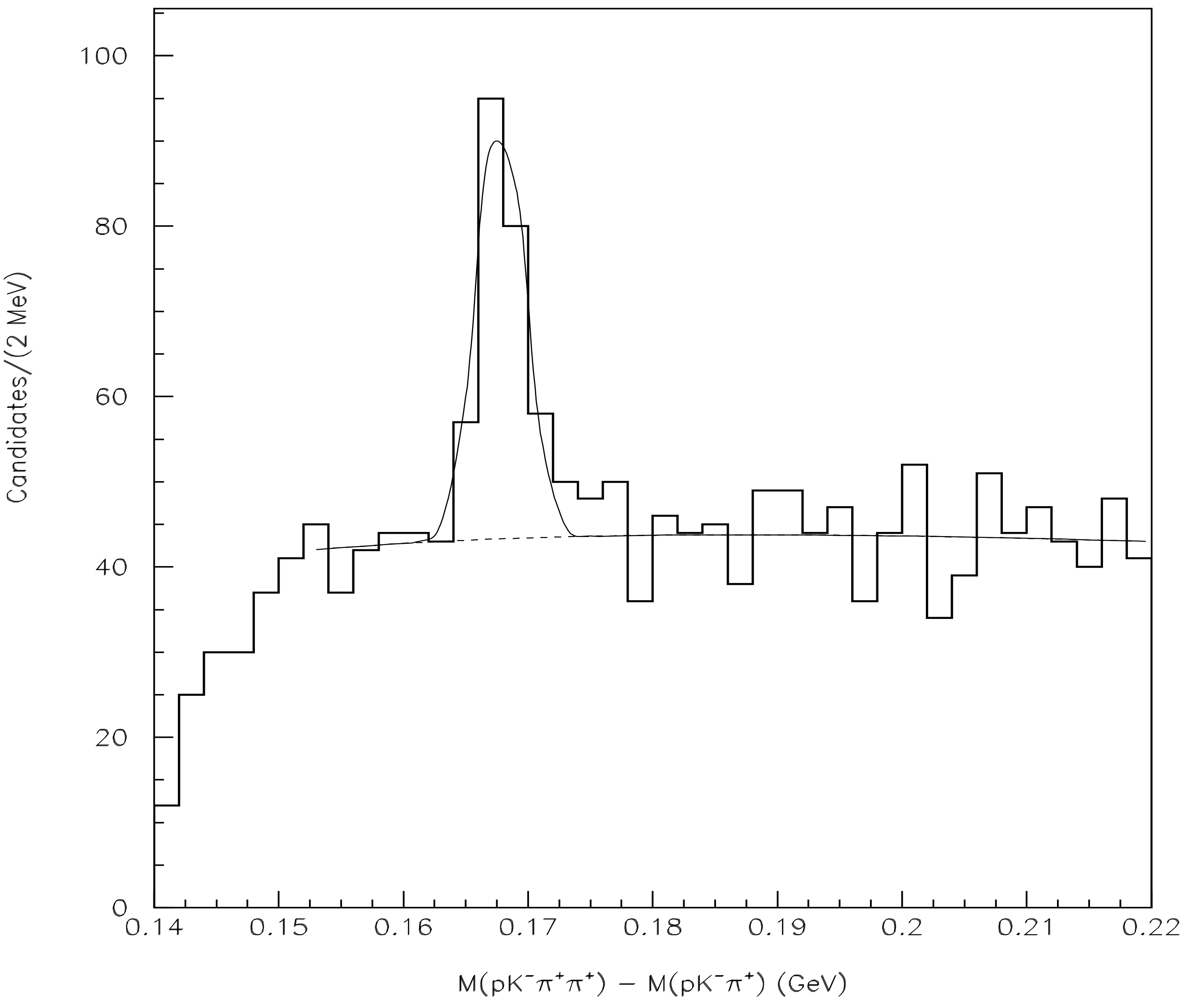

FIG. 3. Mass difference spectrum, M$(pK^-\pi^+\pi^+)$ − M$(pK^-\pi^+)$, for $\Sigma_c^{++} \to pK^-\pi^+\pi^+$ candidates with $pK^-\pi^+$ combinations within $2.5\,\sigma$ of the fitted $\Lambda_c$ mass. The solid line is a fit to a Gaussian distribution plus a background shape described in the text.